\documentclass[useAMS,usenatbib]{mn2e}

\usepackage{graphicx}
\usepackage{subfigure}
 \usepackage{graphicx,amsmath,amssymb}
\usepackage{color} 
\usepackage{hyperref}
\usepackage{color}
\usepackage{amsmath}
\def\lsim{\lower.5ex\hbox{$\; \buildrel < \over \sim \;$}}
\def\gsim{\lower.5ex\hbox{$\; \buildrel > \over \sim \;$}}
\def\simeq{\lower.3ex\hbox{$\; \buildrel \sim \over - \;$}}
\def\ch{\lower-0.55ex\hbox{--}\kern-0.55em{\lower0.15ex\hbox{$h$}}}
\def\lh{\lower-0.55ex\hbox{--}\kern-0.55em{\lower0.15ex\hbox{$\lambda$}}}

\def\etal{{\em et al.} }

\def\etal{{\em et al.}}

\def\lsim{\lower.5ex\hbox{$\; \buildrel < \over \sim \;$}}
\def\gsim{\lower.5ex\hbox{$\; \buildrel > \over \sim \;$}}
\def \simeq{\lower.3ex\hbox{$\; \buildrel \sim \over - \;$}}

\def\Nature{{Nature}}%
\def\araa{{ARA\&A}}%
\def\apj{{ApJ}}%
\def\apjl{{ApJL}}%
\def\apss{{Ap\&SS}}
\def\aap{{A\&A}}%
\def\basi{{BASI}}%
\def\mnras{{MNRAS}}%
\def\na{{New Astron.}}%
\def\nar{{New A Rev.}}%
\def\NCRS{{Nuovo Cimento Rivista Serie}}%
\def\IAUS{{Proceedings IAU Symposium}}%
\def\MmSAI{{Mem. S.A.It}}
\title[Mass loss from advective accretion disc around rotating black holes]
{Mass loss from advective accretion disc around rotating black holes}
\author[Ramiz Aktar, Santabrata Das, Anuj Nandi]
{Ramiz Aktar$^{1}$\thanks{E-mail:
ramiz@iitg.ernet.in (RA); sbdas@iitg.ernet.in (SD), anuj@isac.gov.in (AN)}, 
Santabrata Das$^{1}$, Anuj Nandi$^{2}$
\footnotemark[1] \\ 
$^{1}$Indian Institute of Technology Guwahati, Guwahati, 781039, India.\\
$^{2}$Space Astronomy Group, SSIF/ISITE Campus, ISRO Satellite Centre, Outer
Ring Road, Marathahalli, Bangalore 560037, India}
\begin{document}

\date{Accepted . Received ; in original form }

\pagerange{\pageref{firstpage}--\pageref{lastpage}} \pubyear{}

\maketitle

\label{firstpage}

\begin{abstract}

We examine the properties of the outflowing matter from an advective accretion
disc around a spinning black hole. During accretion, rotating matter experiences
centrifugal pressure supported shock transition that effectively produces a virtual 
barrier around the black hole in the form of post-shock corona (hereafter, PSC).
Due to shock compression, PSC becomes hot and dense that eventually deflects a
part of the inflowing matter as bipolar outflows because of the presence of extra
thermal gradient force. In our approach, we study the outflow properties in terms
of the inflow parameters, namely specific energy (${\mathcal E}$) and specific
angular momentum ($\lambda$) considering the realistic outflow geometry around
the rotating black holes. We find that spin of the black hole ($a_k$) plays an
important role in deciding the outflow rate $R_{\dot m}$ (ratio of mass flux of
outflow and inflow), in particular, $R_{\dot m}$ is directly correlated
with $a_k$ for the same set of inflow parameters. It is found that a large range
of the inflow parameters allows global accretion-ejection solutions and the effective
area of the parameter space (${\mathcal E}$, $\lambda$) with and without outflow
decreases with black hole spin ($a_k$). We compute the maximum outflow rate
($R^{max}_{\dot m}$) as function of black hole spin ($a_k$) and observe that
$R^{max}_{\dot m}$ weakly depends on $a_k$ that lies in the range
$\sim 10\%-18\%$ of the inflow rate for the adiabatic index $(\gamma)$ with 
$1.5 \ge \gamma \ge 4/3$. We present the observational implication
of our approach while studying the steady/persistent Jet activities based on the
accretion states of black holes. We discuss that our formalism seems to have
the potential to explain the observed Jet kinetic power for several Galactic
Black Hole sources (GBHs) and Active Galactic Nuclei (AGNs).
\end{abstract}


\section{Introduction}

Powerful Jets and outflows are commonly observed in accreting black hole
systems including active galactic nuclei (AGNs) and X-ray binaries (XRBs)
\citep{Mirabel-etal92,Mirabel-Rodriguez94,Hjellming-Rupen95,Ferrari98,
Mirabel-Rodriguez98,Junor-etal99,Cheung02,Mirabel03,Miller-etal12}. In spite of the
availability of wealth of high resolution observations, the physical
mechanism of Jet generation and its powering processes are still remain
unclear. However, the obvious cause seems to be the extreme gravity that
powers the outflows where spin of the central objects may play an important
role in producing the relativistic Jets. Earlier work of \citet{Penrose69}
demonstrated that infalling particles on to a rotating black hole have the
potential to extract some of the rotational energy of the black holes. Close to
the horizon, space-time geometry is dragged due to black hole rotation and
the magnetic field lines are twisted resulting the transport of energy from
the black hole along the field lines. \citet{Blandford-Znajek77} showed that
this purely electromagnetic energy extraction mechanism via the threading of
magnetic field lines has the potential to power the Jets. This appealing
mechanism clearly indicates that black hole spin perhaps play an important
role to produce Jets in black hole systems. In order to establish this feature
observationally, attempts have been made to find the evidence of spin-powered
Jets for the stellar mass black holes. According to \citet{Steiner-etal13} and
\citet{McClintok-etal14}, significant positive correlation between the radio
luminosity associated with the mechanical power of ballistic Jets and the spin
of the black hole is seen when the transient black hole systems are fed at
Eddington-limited accretion rates. However, \citet{Russell-etal13} and
\citet{Fender-Gallo14} ruled out the direct evidence of any such correlation
between the transient Jet power and the black hole spin. Using radio-optical
correlation, \citet{Velzen-Falcke13} also reported that the black hole spin
possibly does not have any dominant role in powering the Jets. In a numerical
endeavor, \citet{De-Villiers-etal05} claimed that the Jet efficiency increases
by two orders of magnitude for rapidly rotating black hole compared to the
non-spinning black hole. In the similar context, \citet{Fernandez-etal15} showed
that the more mass is evacuated from the disc as the black hole spin
is increased. In an early effort, \citet{Donea-Biermann96} also found that the
Jet power is strongly dependent on the spin of the black holes without
considering the Jet geometry. Overall, all the above findings are in contrast
and therefore, inconclusive.

Meanwhile, some authors pointed out that there exists a confirm connection between the
accretion spectral states and the Jet states (i.e., thick and thin) in black hole systems,
particularly for microquasars \citep{Gallo-etal03,Rushton-etal10}. In the {\it low-hard spectral states}
({\it LHS}) of galactic black holes, quasi-steady and persistent Jets are observed whereas
relatively stronger Jets are
launched generally in the {\it hard-intermediate state} ({\it HIMS}). In addition, transient
relativistic ejections are observed during the state transition from
{\it hard-intermediate} to {\it soft-intermediate state} ({\it SIMS})
\citep{Fender-etal04,Fender-etal09}. 
Interestingly, Jet activities are not seen in the {\it high-soft states} ({\it HSS}).
It has also been observed that during ejections, Quasi-Periodic Oscillations (QPOs) 
are not observed and energy
spectra get softens \citep{Vadawale-etal01,Millerjones-etal12,Radhika-Nandi14}. 
This is conjectured in a way that the matter is ejected from the inner part of the disc
(i.e., Compton corona equivalently PSC)
\citep{Feroci-etal99,Nandi-etal01a,Chakrabarti-etal02}. Towards this, very recently,
\citet{Radhika-Nandi14} reported the inflow-outflow connection through the observation in
{\it Radio} and {\it X-ray} bands for the object XTE J1859+226 and claimed that
the fast variation of the `spectro-temporal' properties
of outbursting GBHs are connected with the disruption of the post-shock
disc (i.e., PSC) in terms of evacuation of
matter in the form of Jets \citep{Radhika-etal15}. They pointed out that the QPOs are not seen during
`transient' ejection of Jets observed in {\it Radio}. This result indicates that
PSC possibly be responsible for the origin of QPOs in GBHs
\citep{Chakrabarti-Manickam00,Nandi-etal01a,Nandi-etal01b}.
When Jets are emerged out at the cost of the evacuation/disruption of PSC, QPOs are
missing \citep{Vadawale-etal01,Radhika-Nandi14}. Extensive MHD simulations
of accretion disk around rotating black holes also indicate
that windy hot materials (i.e., corona) blows away from the inner part of the disc
as Jets \citep{Koide-etal02,McKinney-Gammie04,De-Villiers-etal05}. Meanwhile, \citet{Das-etal14}
showed that PSC modulates quasi-periodically when viscosity parameter
is chosen to its critical value and such modulation successfully exhibits 
quasi-periodic variation of PSC as well as the outflow rates. These scenarios
indicate that there seems to be direct correlation between PSC and outflow.
Based on these correlations, one can conjecture that the disk-jet symbiosis is
strongly coupled and advective accreting disc perhaps be responsible for the
launching of Jets and outflows.

Advective accretion flow around the black holes must be transonic in order to 
satisfy the inner boundary conditions. Inflowing matter experiences a virtual
barrier in the vicinity of the black hole due to centrifugal repulsion against
gravity. According to the second law of thermodynamics, such a virtual barrier
triggers the discontinuous transition of the flow variables in the form of
shock wave in order to prefer the high entropy solution when possible
\citep{Fukue87,Chakrabarti89,Lu-etal99,Becker-Kazanas01,Fukumura-Tsuruta04}.
Due to compression, the shock induced accretion flow produces hot and dense
post-shock corona (PSC) surrounding the black holes which essentially
acts as the effective boundary layer of the black holes. During accretion, a part of
the inflowing matter is deflected by PSC which is further driven out in the
vertical direction by the excess thermal gradient force across the shock, producing
bipolar outflows. Such an appealing mechanism to launch outflow from the vicinity of
the black hole has been confirmed through numerical simulations
\citep{Molteni-etal94,Molteni-etal96,Machida-etal00,De-Villiers-etal05,Okuda14,Das-etal14,Okuda-Das15}.
In addition, numerous attempts have been made theoretically to calculate the
mass outflow rate around black
holes. \citet{Chakrabarti99} calculated the mass outflow rate considering
isothermal flow. Using both Keplerian and sub-Keplerian components,
\citet{Das-etal01a} self-consistently estimated mass loss from the disc and
found that outflow rate strongly depends on both components.
\citet{Singh-Chakrabarti11} implemented the energy dissipation across the shock while
obtaining the outflow rate and found that possibility of mass loss anti-correlates
with the dissipation rate. Following the work of \citet{Molteni-etal96}, 
several authors
\citep{Chattopadhyay-Das07,Das-Chattopadhyay08,Kumar-Chattopadhyay13,Kumar-etal13,Das-etal14}
studied the properties of mass outflow rate in terms of inflow parameters considering 
dissipative accretion flow. However, all these works were carried out
with limitation as spin of the black hole was not considered. 

Motivating with this, we model the inflow-outflow activity around a rotating
black hole. Since the PSC is induced by the shock transition itself, we self-consistently
calculate the properties of shock driven outflows in terms of the inflow parameters
and investigate the effect of the black hole spin on it. For simplicity, we adopt
the pseudo-Kerr potential proposed by \citet{Chakrabarti-Mondal06} to describe the space time
geometry around the black hole. This potential accurately reproduces the particle
trajectories around rotating black holes having spin $a_{k} \lesssim 0.8$. We calculate
the global accretion solutions in presence of thermally driven outflows and obtain
the parameter spaces spanned by energy and angular momentum of the inflowing matter 
in terms of the black hole spin.
We observe that the shock induced global inflow-outflow
solutions exist for wide range of inflow parameters. Varying the inflow parameters,
we estimate the {\it maximum mass outflow rate} ($R^{\rm max}_{\dot m}$) as function of black hole
spin ($a_k$) and find that $R^{\rm max}_{\dot m}$ weakly depends on $a_k$ having highest
value $\sim 17\%-18\%$ of the inflow rate for $\gamma=4/3$. Further, we
calculate the {\it unabsorbed}
X-ray flux of several black hole objects using the {\it RXTE$^1$}
\footnotetext[1]{http://www.heasarc.gsfc.nasa.gov} satellite archival data and obtain their accretion
rates considering the accretion efficiency $\eta=0.3$ \citep{Thorne74}, which perhaps
reasonable for rotating black holes. With this, using our theoretical estimate of maximum 
mass outflow rate, we then compute the maximum Jet kinetic
power ($L^{\rm max}_{Jet}$). These results are further compared with those available from
observations and close agreements are seen. Following this, our theoretical
prediction essentially provides an estimate of $L^{\rm max}_{Jet}$ for those sources for which
observed Jet power is uncertain.

In the next Section, we describe the basic assumptions and the governing equations
for our model. In \S 3, we discuss the methodology to calculate the mass outflow
rates self-consistently. In \S 4, we present the results which is followed by
discussions. In \S 5, we apply our formalism to calculate the Jet kinetic power
for several astrophysical objects (GBHs and AGNs). Finally in \S 6, we draw the
concluding remarks.

\section{Model Equations and Assumptions}

We consider an axisymmetric disc-jet system
around a rotating black hole in the steady state. Here, the accretion
disc lies along the black hole equatorial plane while the Jet geometry
is described about the black hole rotation axis. In this work, we mainly
focus on the inner region of the accretion disc, where the viscous time scale
is larger than the in-fall time scale and therefore, the angular momentum
transport due to the differential motion is weakly significant leaving the
flow to be inviscid there \citep{Chakrabarti89}. 
Further, in our model, Jets are originated from the inner part of the accretion disc (PSC)
with the same angular momentum of the disc as we neglect the effect of resulting torque
in the disc-jet system. Next, we present the governing equations that
describe the fluid properties of the accretion disc and the Jets. All the
equations are written in geometric unit system as $G=M_{BH}=c=1$, where,
$G$ is the Gravitational constant, $M_{BH}$ is the black hole mass and $c$
is the speed of light, respectively. In this system, the unit of length, mass
and time are expressed as $GM_{BH}/c^2$, $M_{BH}$ and $GM_{BH}/c^3$.

\subsection{Governing equations for Accretion}

We consider a geometrically thin, axisymmetric, low angular momentum,
advective accretion flow around a rotating black hole. For simplicity, we
adopt the pseudo-Kerr effective potential introduced by \citet{Chakrabarti-Mondal06} to
represent the space-time geometry around the black hole
instead of using full general relativistic prescription. This enables us
to solve the problem following the Newtonian approach and keeping all
the salient features of space-time geometry around it.
The equation of motion describing the accreting matter is given by,

(i) the energy conservation equation:

$$
\mathcal{E}=\frac{v^{2}}{2}+\frac{a^{2}}{\gamma -1}+\Phi  ,
\eqno(1)
$$ 
where, $\mathcal{E}$ represent the specific energy of the flow,
$v$ is the radial velocity and $a$ is the adiabatic sound speed 
defined as $a=\sqrt{\gamma P/\rho}$. Here, $P$ is the isotropic
pressure, $\rho$ is the gas density, $\gamma$ is the adiabatic index, respectively.
The accreting gas is described by the adiabatic equation of state as 
$P=K\rho^{\gamma}$, where $K$ is the measure of specific entropy, which
is constant except at the shock transition. The effective potential $\Phi$
is given by \citep{Chakrabarti-Mondal06},

$$
\Phi=-\frac{B+\sqrt{B^2-4AC}}{2A},
\eqno(2)
$$
where,
$$
A=\frac{\alpha^2 \lambda^2}{2x^2},
$$

$$
B=-1 + \frac{\alpha^2 \omega \lambda r^2}{x^2} 
+\frac{2a_k\lambda}{r^2 x},
$$

$$
C=1-\frac{1}{r-x_0}+\frac{2a_k\omega}{x}
+\frac{\alpha^2 \omega^2 r^4}{2x^2}.
$$
Here, $x$ and $r$ represent the cylindrical and spherical radial distance
considering the black hole is located at the origin of the coordinate system
and $\lambda$ is the specific angular momentum of the flow.
Here, $x_0=0.04+0.97a_k+0.085a_k^2$, $\omega=2a_k/(x^3+a^2_k x+2a^2_k)$
and $\alpha^2=(x^2-2x+a^2_k)/(x^2+a_k^2+2a^2_k/x)$, $\alpha$ is the
redshift factor and $a_k$ represents the black hole rotation parameter
defined as the specific spin angular momentum of the black hole. 
According to \citet{Chakrabarti-Mondal06}, the above potential mimic the Kerr geometry quite
satisfactorily for a wide range of $a_k\lesssim0.8$.

(ii) the mass conservation equation:

$$
\dot{M}=4 \pi \rho v x h ,
\eqno(3)
$$
where, ${\dot M}$ denotes the mass accretion rate which is constant everywhere
except the region of mass loss and $h$ is the half-thickness of
the disk obtained from thin disk approximation \citep{Chakrabarti89} as,

$$
h(x)=a\sqrt{\frac{x}{\gamma \Phi_{r}^{'}}},
\eqno(4)
$$
where, $\Phi_{r}^{'}=(\partial{\Phi}/\partial{r})_{z<<x}$, $z$ is the vertical height in
the cylindrical coordinate system and $r=\sqrt{x^{2}+z^{2}}$.

Combining the expression of sound speed and the adiabatic equation of the state of
the gas, we calculate the entropy accretion rate as \citep{Chakrabarti90},

$$
\dot{\mathcal{M}}=v a^{\nu} \sqrt{\frac{x^{3}}{\gamma \Phi_{r}^{'}}},
\eqno(5)
$$
where, $\nu$ = $(\gamma+1)/(\gamma-1)$. In an accretion disk, $\dot{\mathcal{M}}$
remain constant all throughout except at the shock transition where local turbulence
generates entropy. Therefore, the entropy accretion rate in the post-shock
region is higher than that in the pre-shock region. 

By definition, the accretion flow around the black holes must be transonic in nature.
This is due to the fact that the velocity of the accreting matter at large distances
from the black hole is negligibly small and therefore, the flow is subsonic. However,
flow crosses the event horizon with velocity equivalent to the speed of light implying
that the flow is supersonic close to the black hole. This clearly indicates that the
accretion flow necessarily
changes its sonic state during its journey from the outer edge of the disc to the horizon.
In order to calculate the location where the state of this sonic transition occurs,
we derive the sonic point conditions using Eqs. (1-5) and obtain as,

$$
\frac{dv}{dx}=\frac{N}{D},
\eqno(6)
$$
where, 
$$
N=\frac{a^{2}}{\gamma+1}\left[\frac{3}{x}-
\frac{d{\rm ln}\Phi_{r}^{'}}{dx}\right]-\frac{d\Phi_{e}}{dx},
\eqno(7)
$$
and
$$
D=v-\frac{2a^{2}}
{(\gamma+1) v},
\eqno(8)
$$
where, the subscript `e' signifies the quantity measured at the disc equatorial plane.
Since the accretion flow is smooth everywhere, the radial velocity gradient must be
finite everywhere too. According to Eq. (6), if the denominator $D$ vanishes at any radial
distance, the numerator $N$ also vanishes there. 
Such a location, where $D=N=0$, is called as sonic point. 
Setting $D=0$ and $N=0$ independently, we find the sonic point conditions as,

$$
D=0 \Rightarrow M_c = \frac{v_c}{a_c} =\sqrt{\frac{2}{\gamma + 1}},
\eqno(9)
$$
and
$$
N=0 \Rightarrow a_c^{2}=(\gamma+1)\left(\frac{d\Phi_e}{dx}\right)_{c}
\left[ \frac{3}{x}-\frac{d{\rm ln}\Phi_{r}^{'}}{dx} \right]_{c}^{-1}.
\eqno(10)
$$
Here, the subscript `c' denotes the quantities evaluated at the sonic points.
We use sonic point conditions, namely, Eqs. (9-10), in Eq. (1) to obtain
the location of sonic point for a given
set of ($\mathcal{E}$, $\lambda$) of the flow. For a physically acceptable
transonic solution, flow must contain at least one saddle type sonic point
(\citet{Das07} and references therein). Depending on the values of $\mathcal{E}$
and $\lambda$, flow may possess multiple sonic points as well which is one of the
necessary condition to form a shock wave \citep{Chakrabarti90}. 
At the sonic point, Eq. (6) takes the form as $dv/dr=0/0$ and therefore, we
use l{'}Hospital rule to calculate the radial velocity gradient there.
Once the flow variables at the sonic point are known, we integrate Eq. (6)
starting from the sonic point inward up to the black hole horizon and outward
to a large distance to obtain the full set of global accretion solution which
may own standing shock waves. 

\subsection{Governing equations for Outflow}

In this work, we consider the outflow to be emerged out from the accretion disc
along the rotation axis of the black hole with the same energy and angular
momentum as the accretion flow since we neglect the dissipative processes (i.e.,
viscosity, cooling etc.). Similar to accretion flow, we consider the
outflow to obey the polytropic equation of state as $P_j=K_j \rho_j^\gamma$,
where, the suffix `$j$' denotes the outflow variables.
The equations of motion that describe the outflow dynamics are given by,

(i) \emph{the energy conservation equation of outflow}:

$$
\mathcal{E}_{j}=\frac{1}{2}v_j^2+\frac{a_j^2}{\gamma -1 }+\Phi,
\eqno(11)
$$
where $\mathcal{E}_{j}$ ($\equiv {\mathcal E}$) represent the specific energy of
the outflow, $v_j$ is the outflow velocity and $a_j$ is the sound speed of the
outflow, respectively.

(ii) \emph{Mass conservation equation of outflow}:

$$
{\dot{M}_{out}}=\rho_{j}v_{j}\mathcal{A},
\eqno(12)
$$
where, $\mathcal{A}$ is a geometrical quantity representing the total area function
of the outflow. To obtain $\mathcal{A}$, we consider the outflow geometry as 
described in \citet{Molteni-etal96} where the outflowing matter tends to come
out through the two surfaces, namely the centrifugal barrier (CB) and the 
funnel wall (FW). 
We calculate the centrifugal barrier by identifying the pressure
maxima surface as $(d\Phi/dx)_{r_{CB}}=0$ and the funnel wall by defining the
null effective potential as $\Phi\vert_{r_{FW}}=0$ \citep{Molteni-etal96}.
As the effective potential of our interest is complex in nature, it is
unattainable to prevail an analytical expression of the CB surface and the FW
and therefore, we compute them numerically.
In Figure 1, we illustrate the outflow geometry around the rotating black holes
for a wide range of $a_k$ marked in the figure and compare them with the same
obtained using the pseudo-Newtonian potential introduced by \citet{Paczynski-Wiita80}
appropriate for a stationary black hole. The regions bounded by the dashed and dotted curves are for
$a_k=-0.998$ and $a_k=0.8$, respectively and the solid curves depict the
result of stationary black hole. Note that the outflow geometry for stationary
as well as for rotating black holes is indistinguishable outside the range
of few tens of Schwarzschild radius. This is possibly due to the fact that the
effect of black hole spin on the space-time geometry cursorily diminishes with
the increasing distances. Keeping this
in mind, we therefore adopt the outflow geometry of the stationary black hole
for our present study in order to avoid the rigorous numerical calculations. This
enables us to obtain an analytical expression of area function $\mathcal{A}$
and its higher order derivatives, which is required in the sonic
point analysis of outflowing matter \citep{Das-Chattopadhyay08}. A comprehensive study of sonic point 
analysis for outflows including the area function $\mathcal{A}$ and its
derivatives have already been presented in \citet{Das-Chattopadhyay08}
and therefore, we avoid repetition here.

%%%%%%%%%%%%%%%%%%%%%%%%%%%%%%%%%%%%%%%%%%%%%%%%%%
\begin{figure}
\begin{center}
\includegraphics[width=0.45\textwidth]{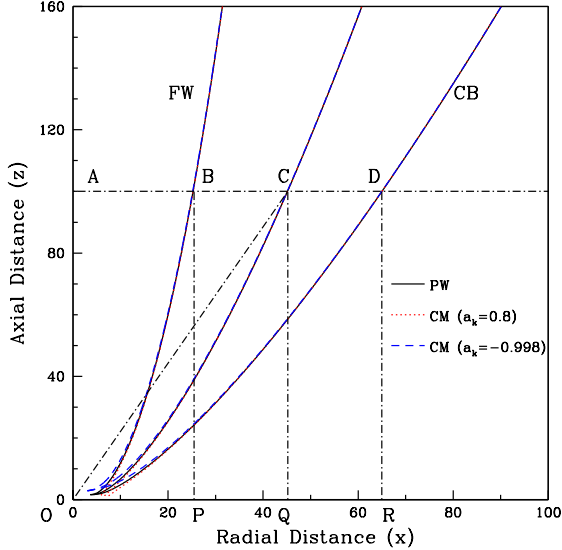}
\end{center}
\caption{Comparison of Jet geometry for angular momentum $\lambda = 3.3$. Funnel wall (FW) and 
centrifugal barrier (CB) are marked in the figure. Here, ${\rm OC}=r_j$ is
the spherical radius representing the streamline of the outflow. Dashed and
dotted are obtained for $a_k=-0.998$ and $0.8$ while solid curve denotes the
result obtained using pseudo-Newtonian potential \citep{Paczynski-Wiita80}. Jet geometries
for stationary as well as rotating black holes are indistinguishable beyond
few tens of Schwarzschild radius. See text for details.
}

\end{figure}
%%%%%%%%%%%%%%%%%%%%%%%%%%%%%%%%%%%%%%%%%%%%%%%%%%

\section{Solution Methodology}

In this study, we consider the post-shock matter (PSC) as the precursor of the Jet base
and thus, we focus only to those accretion solutions that possess standing shock waves. 
For shock, the accretion flow variables experience discontinuous transition characterized by the
Rankine-Hugoniot (hereafter R-H) shock conditions \citep{Landau-Lifshitz59}.
These conditions include the conservations of 
mass flux, energy flux and momentum flux across the shock, respectively.
In presence of mass loss, a part of the inflowing matter is effectively emerged out as
outflow from the post-shock region (PSC), while the remaining matter is advected
in to the black hole straight away.
Therefore, in the present scenario, the shock conditions are given by,

(i) \emph{Conservation of mass flux}:
$$
\dot{M}_{+}=\dot{M}_{-}-\dot{M}_{out}=\dot{M}_{-}(1 - R_{\dot{m}}).
\eqno(13a)
$$
The quantities having subscripts `-' and `+' are referred to the values
before and after the shock which we follow throughout the paper unless
otherwise stated. The pre-shock and post-shock accretion rates are
denoted by $\dot{M}_{-}$ and $\dot{M}_{+}$, respectively and $\dot{M}_{out}$
is the mass flux for the outflowing matter. Following this, the mass outflow
rate is computed as $R_{\dot{m}} = \dot{M}_{out}/\dot{M}_{-}$. The next condition is

(ii) \emph{Conservation of energy flux}:
$$
\mathcal{E_+}=\mathcal{E_-},
\eqno(13b)
$$
and finally, we have

(iii) \emph{Conservation of momentum flux}:

$$
W_{+} + \Sigma_{+}v_{+}^2 = W_{-} + \Sigma_{-} v_{-}^2,
\eqno(13c)
$$
where, $W$ and $\Sigma$ represent the vertically integrated  pressure and 
density \citep{Das-etal01b}. 

We rewrite Eq. (13b) and Eq. (13c) in terms of Mach number
($M=v/a$) and are given by,
$$
\frac{1}{2}M_{+}^{2}a_{+}^{2}+\frac{a_{+}^{2}}{\gamma-1}
=\frac{1}{2}M_{-}^{2}a_{-}^{2}+\frac{a_{-}^{2}}{\gamma-1} 
\eqno(13d)
$$
and
$$
\left[\frac{2a_{+}}{(3\gamma-1)M_{+}}+M_{+}a_{+}\right] =
\frac{1}{1-R_{\dot m}}\left[\frac{2a_{-}}{(3\gamma-1)M_{-}}+M_{-}a_{-}\right],
\eqno(13e)
$$
where we utilize Eq. (13a) and $a=\sqrt{(3\gamma-1)W/(2\Sigma)}$ \citep{Das-etal01b}.

Using Eqs. (13d-e), we obtain a shock invariant 
quantity ($C_{s}$) in terms of Mach number $M$ as,

\begin{align*}
C_{s} = \frac
{\left[\frac{2}{M_{+}}+(3\gamma-1)M_{+}\right]^{2}(1-R_{\dot{m}})^{2}}{\left[2+(\gamma-1)M_{+}^{2}\right]}
= \frac
{\left[\frac{2}{M_{-}}+(3\gamma-1)M_{-}\right]^{2}}{\left[2+(\gamma-1)M_{-}^{2}\right]},
\end{align*}
where, $M_{-}$ and $M_{+}$ stand for Mach number just before and after the
shock in the pre-shock and post-shock flow, respectively.

Since the shock conditions are coupled with inflow and outflow variables,
the accretion and Jet equations are solved simultaneously. We obtain the Jet
velocity gradient from its governing equations and calculate
the sonic point properties following the `critical point' analysis \citep{Chakrabarti89}.
While doing this, we use the inflow parameters, namely, energy 
$\cal E$ and angular momentum $\lambda$ of the inflow. This is because
the outflow is considered to be originated from the the post-shock region (PSC). This
also ensures that the Jets are launched with the same density as the post-shock
flow, namely $\rho_{j} = \rho_{+}$. Further, we integrate the
Jet equations to calculate the outflow variables
at the Jet base starting from the Jet sonic point and using
Eqs. (3), (4) and (12) we compute the mass outflow rate  in terms of the inflow-outflow
properties at the shock and is given by,
$$
R_{\dot{m}}=\frac{\dot{M}_{out}}{\dot{M}_{-}}
=\frac{\rho_{j} v_{j}(x_{s}) \mathcal{A}(x_{s})}{4\pi \rho_{-} v_{-} x_{s} h_{-}}
=\frac{Rv_{j}(x_{s})\mathcal{A}(x_{s})}
{4\pi \sqrt{\frac{1}{\gamma}} x_{s}^{3/2}{\Phi_{r}^{'}}^{-1/2}a_{+}v_{-}},
\eqno(14)
$$
where, %$R=\Sigma_{+}/\Sigma_{-}$, 
$R=\Sigma_{+}/\Sigma_{-} (\equiv \rho_{+} h_{+}/\rho_{-}h_{-}
= \rho_{j} a_{+}/\rho_{-} a_{-})$,
is the compression ratio and $v_{j}(x_{s})$ 
and $\mathcal{A}(x_{s})$ are the Jet velocity and the Jet area function at the
shock, respectively. We utilize an iterative method to calculate $R_{\dot{m}}$
self-consistently which is given as follows:  

\noindent We begin with $R_{\dot{m}}=0$. Using the shock invariant quantity, we calculate
the virtual shock location $x^{\rm v}_{s}$ for a given set of ($\cal E$, $\lambda$) with
the consideration that the entropy of the inflowing post-shock matter
($\dot {\cal M}_{+}$) and outflowing matter ($\dot {\cal M}_{out}$) are larger than the
entropy of the pre-shock matter ($\dot {\cal M}_{-}$). This eventually reflects the 
fact of second law of thermodynamics as the shocked solutions ascertain the
preferred mode of accretion. We use the same set of inflow parameters
for Jet equations and calculate $R_{\dot{m}}$ which we employ further in the shock 
invariant equation to obtain a new shock location. We continue this iteration
process until the solution converges to the actual shock location and accordingly
we compute the corresponding $R_{\dot{m}}$. In the following Sections, we investigate
the properties of $x_s$ and $R_{\dot{m}}$ in terms of the inflow parameters 
($\mathcal {E}$, $\lambda$) for various values of black hole spin ($a_k$) and
present the results which are followed by the discussion on astrophysical applications.

\section{Results and Discussions}

In an advective accretion process around black holes, the accreting matter
suffers discontinuous shock transitions due to the centrifugal barrier. This
causes the post-shock flow to become hot and puffed up which eventually
behaves like a effective boundary layer around the black hole (i.e., PSC). During accretion,
a part of the inflowing matter is deflected vertically by this boundary
layer to produce thermally driven outflows. To illustrate the complete picture
of the accretion and ejection mechanism, we present the global inflow-outflow
solutions for various $a_k$ in Fig. 2. In panel (a), (b) and (c), we show the
variation of Mach number ($M=v/a$) of the inflow with the radial distance ($x$) for
$a_k$ = 0.8, 0.0, and -0.998, respectively. Subsonic matter at large distance
gradually gains its radial velocity as it proceeds towards the black hole due
to the fatal attraction of gravity and crosses the outer sonic point ($x_{out}$) to
become supersonic as shown by the arrows. As the flow continues its journey
further towards the black hole, it experiences discontinuous transition in
flow variables when the R-H conditions are favorable which is indicated by
the dashed vertical arrow. After this transition, flow momentarily slows down and
gradually picks up its velocity. Eventually, flow becomes supersonic again
after crossing the inner sonic point ($x_{in}$) and finally enters in to the
black hole. As the outflow is emerged out from the effective boundary layer
of the black hole, the density of the post-shock flow is decreased causing the
reduction of pressure at PSC as well. In order to maintain the pressure balance
across the standing shock in presence of mass loss, shock itself has to move
inward which is indicated by the solid vertical arrow. In the panel (d),
we plot the Mach number ($M_j$) variation of the outflowing matter with its radial
coordinate ($x_j$) corresponding to panel (a), (b) and (c). The black solid
circles represent the sonic locations and the arrows show the direction of motion
of the outflowing matter. The inflow parameters for panel (a) is (${\cal E},~\lambda$)
=(0.007, 2.65), for (b) is (0.0025, 3.45) and for (c) is (0.0015, 4.0), respectively. 

%%%%%%%%%%%%%%%%%%%%%%%%%%%%%%%%%%%%%%%%%%%%%%%%%%
\begin{figure}
\begin{center}
\includegraphics[width=0.5\textwidth]{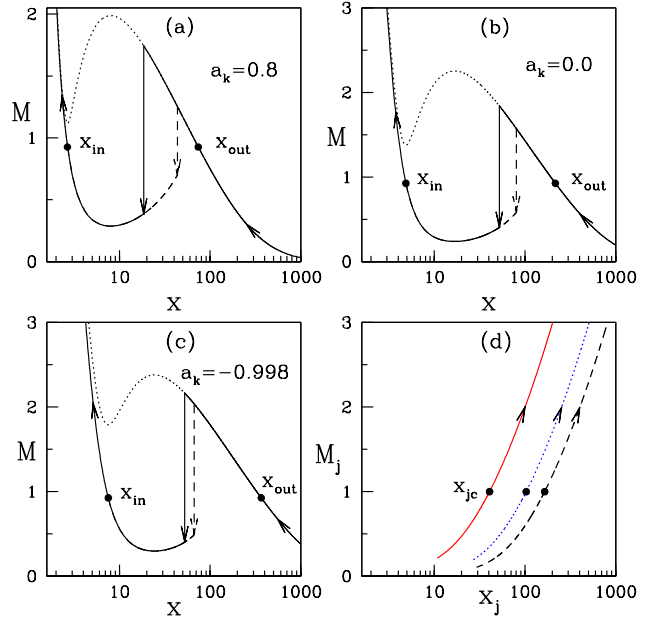}
\end{center}
\caption{ 
Variation of the inflowing Mach number $M$ $(=v/a)$ with radial distances ($x$) for 
(a) $a_k=0.8$, (b) $a_k=0$ and (c) $a_k=-0.998$, respectively. The corresponding
outflow Mach number $M_j$($=v_j/a_j$) variation is shown in panel (d). See text
for details. 
}

\end{figure}
%%%%%%%%%%%%%%%%%%%%%%%%%%%%%%%%%%%%%%%%%%%%%%%%%%

%%%%%%%%%%%%%%%%%%%%%%%%%%%%%%%%%%%%%%%%%%%%%%%%%%
\begin{figure} 
\includegraphics[width=0.48\textwidth]{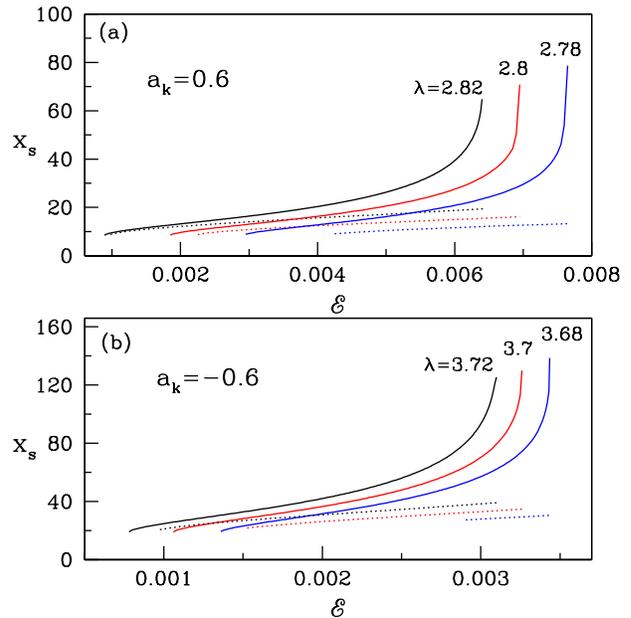}
\caption{ Variation of shock location as function of energy $\mathcal E$.
The solid curves represent results without mass loss and the dotted curves
are with mass loss. In the upper panel, we choose $a_k=0.6$ and curves are
for $\lambda=2.78$ (right), $2.80$ (middle) and $2.82$ (left), respectively.
In the lower panel, we consider $a_k=-0.6$ and curves are
for $\lambda=3.68$ (right), $3.70$ (middle) and $3.72$ (left), respectively.
}
\end{figure}
%%%%%%%%%%%%%%%%%%%%%%%%%%%%%%%%%%%%%%%%%%%%%%%%%%

In Fig. 3, we compare the location of the shock transitions as function of
energy $\mathcal E$ for a set of angular momentum $\lambda$. In the upper panel,
we choose $a_k=0.6$ and find that the inflow-outflow solution possesses shock wave
for a wide range of $\mathcal E$ and $\lambda$. As anticipated in Fig. 2, the shock
location moves towards to the black hole horizon when $R_{\dot m}\ne0$ just to
maintain the pressure balance across the shock. We again observe that the shock location
reaches to its lowest value around $x^{\rm min}_s \sim 8r_g$ at energies higher compared
to the case of $R_{\dot m}=0$ and the limiting value of this energy increases with
the decrease of angular momentum of the flow. However, the maximum value of energy
that allows shock transition in presence of mass loss is identical with no mass loss case.
This provides a clear hint that the range of inflow parameters for outflows, namely
energy $\mathcal E$ and angular momentum $\lambda$ are reduced from their lower ends. 
In the lower panel, we consider $a_k=-0.6$ and obtained the similar results which
differ only quantitatively. Overall, it is clear that the possibility of shock formation
is affected substantially due to the presence of outflow.

%%%%%%%%%%%%%%%%%%%%%%%%%%%%%%%%%%%%%%%%%%%%%%%%%%
\begin{figure}
\includegraphics[width=0.5\textwidth]{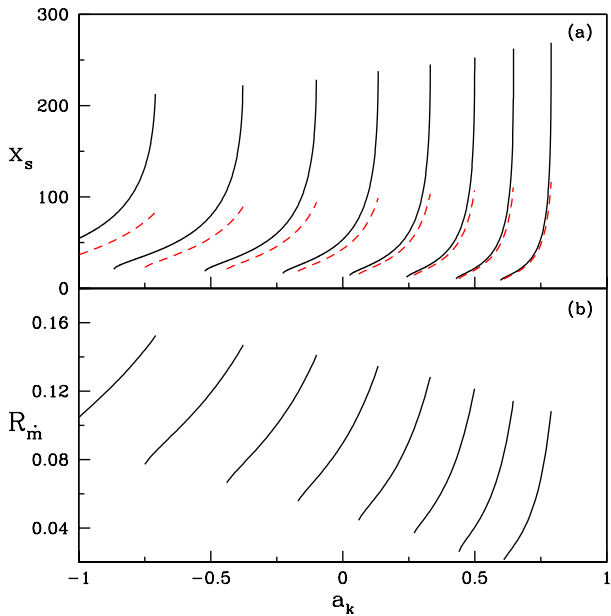}
\caption{Variation of (a) shock location $(x_{s})$ and (b) outflow rate
$(R_{\dot{m}})$ as function of black hole rotation parameter $a_{k}$ for
$\lambda$ = 2.8 to 3.92 (right to left), where $\Delta \lambda =0.16$.
Here, we fixed the flow energy as $\mathcal{E}$ = 0.002. In the upper
panel, dashed curves denotes the variation of shock location in presence
of outflow.
}
\end{figure}
%%%%%%%%%%%%%%%%%%%%%%%%%%%%%%%%%%%%%%%%%%%%%%%%%%

Before we proceed further, we now investigate the effect of the black hole rotation
on the generation of mass outflow rate. In Fig. 4, we present the variation of
shock location ($x_{s}$) and outflow rate ($R_{\dot{m}}$) as function of $a_{k}$.
Here, we fix $\mathcal{E} = 0.002$ and vary $\lambda$ from $2.8$ to 3.92 from 
the right most curve to the left most with an interval $\Delta \lambda =0.16$.
In the upper panel, the solid curves represent the shock locations in absence
of mass loss. However, in presence of mass loss, accretion flow adjusts the
location of the shock transition closer to the black hole horizon 
in a way that the pressure balance condition across the shock is maintained
which is depicted by the dashed curves. On the contrary, as $a_k$ is increased
for a set of $\mathcal{E}$ and $\lambda$,
the shock locations recede away from the black hole horizon. 
This indicates that the size of the post-shock region (PSC) is enhanced with the
increase of $a_k$ and the inflowing matter is eventually intercepted by the
large effective area of the post-shock flow that produces 
more outflow rate. In the lower panel, we present the feature of outflow
rate variation with $a_k$. For a set of $\mathcal{E}$ and $\lambda$, the
outflow rate $R_{\dot{m}}$ shows non-linear correlation for prograde as well as
retrograde flows. In addition, we observe that for a given $a_k$, $R_{\dot{m}}$
is higher for increasing $\lambda$ and with this, we infer that large
outflow rate is associated with the higher $\lambda$ and lower $a_k$ when 
${\mathcal E}$ is fixed. Overall, we find that outflow rate reaches close to 16\% for
these parameters as depicted in the figure.

%%%%%%%%%%%%%%%%%%%%%%%%%%%%%%%%%%%%%%%%%%%%%%%%%%
\begin{figure} 
\includegraphics[width=0.5\textwidth]{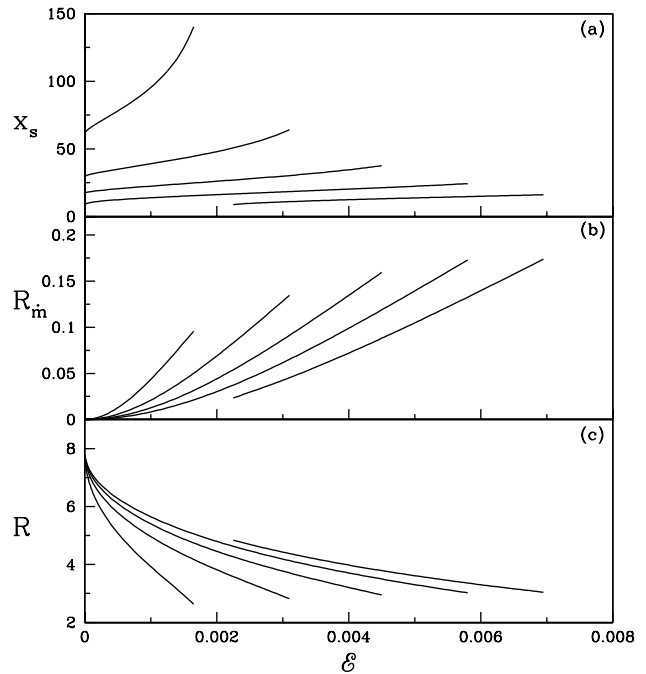}
\caption{ Variation of (a) shock location $(x_{s})$,  (b) outflow rate 
$(R_{\dot{m}})$ and (c) compression ratio $(R)$ as function of energy
$\mathcal E$ for $a_{k} = 0.6, 0.65, 0.7, 0.75$ and $0.8$ (right to left).
Here, we fix the flow angular momentum $\lambda = 2.8$. }
\end{figure}

%%%%%%%%%%%%%%%%%%%%%%%%%%%%%%%%%%%%%%%%%%%%%%%%%%

Next, we study the characteristics of the post-shock quantities in terms
of the inflow variables and present them in Fig. 5. Here, we choose the angular
momentum as $\lambda = 2.8$ and vary energy $\mathcal E$ and black hole rotation
parameter $a_k$. In the upper panel, we plot the variation of shock location
($x_s$) with $\mathcal E$ for $a_k$ varied from $0.6$ (right) to 0.8 (left)
with $\Delta $$a_{k}=0.05$. The corresponding variation of outflow rate
$R_{\dot{m}}$ is shown in the middle panel. Due to shock transition, the
post-shock flow is compressed. The measure of this compression is 
quantified as the ratio of the post-shock density to the pre-shock density
and it is termed as shock compression ratio $R$. In the lower panel, we present
the variation of $R$ with $\mathcal E$. When shock forms closer to the
black hole horizon, the amount of gravitational potential energy release is
higher ensuing the formation of strong shock. As the energy is increased, 
the shock location moves outward that increases the outflow rate and weakens
the shock compression ratio. We observe similar behaviour for all $a_k$.
However, for a given $\lambda$ and $a_k$, there is a range of energy beyond
which mass outflow ceases to exist. This provides an indication of finding 
a parameter space for shock in the fundamental plane of energy and angular momentum
of the inflow. We perform the similar study for retrograde flow as well 
and find the identical trend as depicted in Fig. 6.

%%%%%%%%%%%%%%%%%%%%%%%%%%%%%%%%%%%%%%%%%%%%%%%%%%
\begin{figure}
\begin{center}
\includegraphics[width=0.5\textwidth]{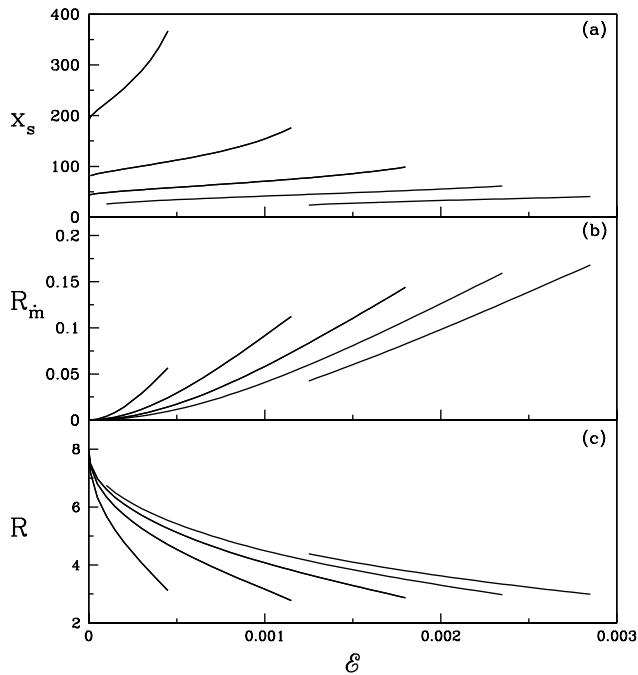}
\end{center}
\caption{Same as Fig. 5. Here, $a_{k}= -0.998$, $-0.8$, $-0.6$,
$-0.4$, and $-0.2$ (right to left) and flow angular momentum is 
$\lambda = 3.9$.
}
\end{figure}
%%%%%%%%%%%%%%%%%%%%%%%%%%%%%%%%%%%%%%%%%%%%%%%%%%

%%%%%%%%%%%%%%%%%%%%%%%%%%%%%%%%%%%%%%%%%%%%%%%%%%
\begin{figure}
\includegraphics[width=0.45\textwidth]{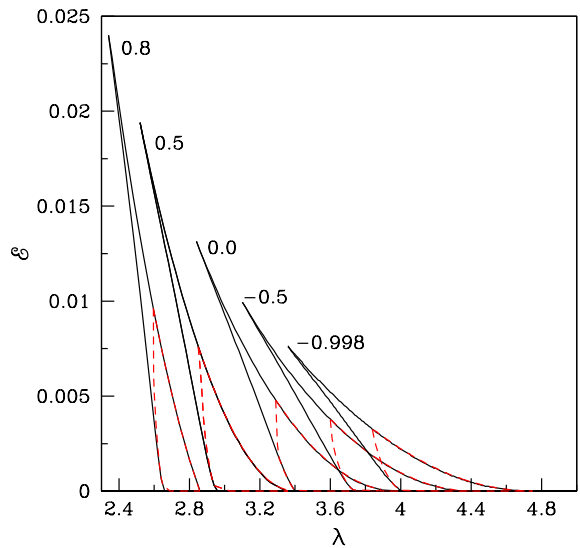}
\caption{Energy angular momentum parameter space. Solid curve (black): shock 
parameter space without outflow. Dashed curve (red): shock parameter space 
with outflow. Various values of $a_k$ are marked in the figure. }
\end{figure}

%%%%%%%%%%%%%%%%%%%%%%%%%%%%%%%%%%%%%%%%%%%%%%%%%%

As we pointed out earlier that the transonic accretion solutions including
R-H shock waves are not isolated solutions. In fact, these solutions do exist
for a wide range of parameters, namely, energy ($\mathcal E$), angular momentum
($\lambda$), and black hole rotation parameter ($a_k$). Towards this, we
identify the shock induced global accretion solutions using the parameter space
spanned by the energy ($\mathcal E$) and angular momentum ($\lambda$) of the
flow and classify them in terms of $a_k$. In Fig. 7, we separate the parameter
spaces with the solid boundaries that indicate the regions for global shock
accretion solutions in absence of mass loss. The corresponding values of $a_k$
are marked in the figure. We further investigate the parameter spaces that 
cater mass loss and indicate it with the dashed boundaries. Due to mass loss,
the post-shock region (PSC) shrinks as seen in Fig. 2, and therefore, the associated
parameter space is reduced compared to the results having no outflows, particularly
towards the lower energy and lower angular momentum sides (see Fig. 3).
For flows with input parameters chosen from these part of the parameter space,
shock forms very close to the black hole horizon when outflow
is ignored \citep{Das-etal01b} and they cease to exist when the outflow is 
allowed to emerge out from the post-shock disc.

%%%%%%%%%%%%%%%%%%%%%%%%%%%%%%%%%%%%%%%%%%%%%%%%%%
\begin{figure}
\includegraphics[width=0.45\textwidth]{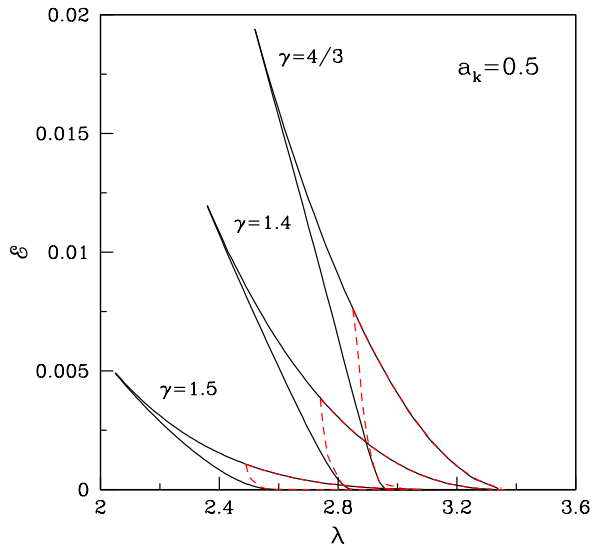}
\caption{Effective regions of parameter space separated in terms of shock waves
transition for various values of $\gamma$ marked in the figure. Here, we choose $a_{k}=0.5$.
The solid boundary denotes the presence of shock waves
in absence of mass loss and the dashed boundary represents the results including
mass loss.
}
\end{figure}
%%%%%%%%%%%%%%%%%%%%%%%%%%%%%%%%%%%%%%%%%%%%%%%%%%

So far, we have presented the shock induced complete global inflow-outflow solutions
and its properties for a given value of $\gamma=4/3$. In reality, however, the theoretical
limit of $\gamma$ lies in the range between $4/3$ to $5/3$ depending on the ratio between
the thermal energy and the rest energy of the flow \citep{Frank-etal02}. To infer this,
we consider rotating flows that are characterized as thermally ultra-relativistic
($\gamma \sim 4/3$), thermally trans-relativistic ($\gamma \sim 1.4$) and thermally
semi-non-relativistic ($\gamma \sim 1.5$), respectively \citep{Kumar-etal13} and
obtain the parameter spaces for $a_k = 0.5$ similar to Fig. 7. This is shown in Fig. 8.
Here, we observe that
in all three cases shocks exist for significant range of inflow parameters.
However, the range of parameters for shock reduces considerably as the flow
changes its ultra-relativistic character towards the non-relativistic limit. In
addition, we find that all the parameter spaces further shrink as in Fig. 7, when
a part of the inflowing matter is deflected in the form of mass loss from the inner
part of the disc (PSC).

%%%%%%%%%%%%%%%%%%%%%%%%%%%%%%%%%%%%%%%%%%%%%%%%%%
\begin{figure}
\includegraphics[width=0.45\textwidth]{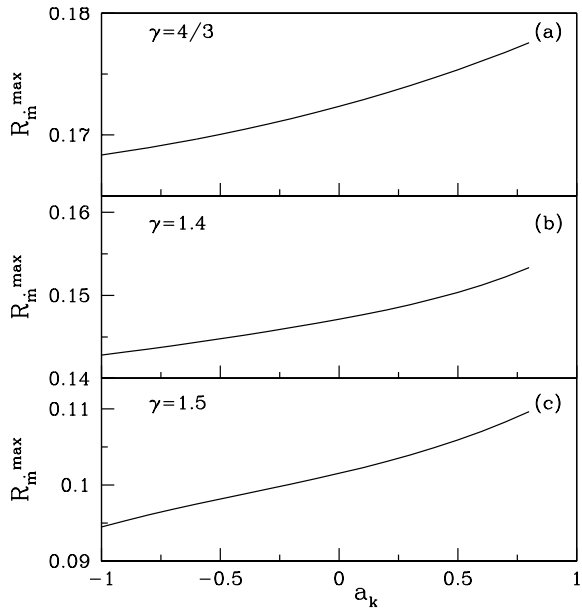}
\caption{Variation of maximum outflow rates $R^{\rm max}_{\dot{m}}$ with the 
black hole rotation parameter $a_{k}$. Upper panel is for $\gamma = 4/3$ (a),
middle panel is for $\gamma = 1.4$ (b) and bottom panel is for $\gamma =1.5$ (c),
respectively. See text for details.
}
\end{figure}
%%%%%%%%%%%%%%%%%%%%%%%%%%%%%%%%%%%%%%%%%%%%%%%%%%

According to the formalism adopted in this work, the mass outflow rate $R_{\dot m}$
is computed self-consistently in terms of the inflow parameters around a
black hole of rotation parameter $a_k$. This allows us to estimate the maximum 
$R^{max}_{\dot m}$ for a given value of $\gamma$. While doing this, we identify a particular
set of energy ($\mathcal E$) and angular momentum ($\lambda$) from their
full range that provides the highest value of $R_{\dot m}$. The importance of this
investigation is associated with the study of the maximum Jet kinetic power
corresponding to $R^{max}_{\dot m}$, which we discuss in the next section.
Following Fig. 8, we calculate the variation of $R^{max}_{\dot m}$ as function
of $a_k$ for various values of $\gamma$ which is presented in Fig. 9.
The results displayed in
upper, middle and bottom panels are for $\gamma = 4/3$, $1.4$ and $1.5$, 
respectively. We find that $R^{max}_{\dot m}$ gradually increase with the
increase of $a_k$ for all cases. 
We also observe that the accretion flows having $\gamma=4/3$ have the potential to
produce more outflows compared to the flows with higher values of $\gamma$. This
is perhaps inevitable due to the fact that the outflows under consideration
are thermally driven and therefore, thermally ultra-relativistic flows exhibit
maximum outflows. Following this, in the next Section, we consider $\gamma=4/3$
while computing the Jet kinetic power for GBHs and AGNs, until otherwise stated.

%%%%%%%%%%%%%%%%%%%%%%%%%%%%%%%%%%%%%%%%%%%%%%%%%%
\begin{figure}
\includegraphics[width=0.45\textwidth]{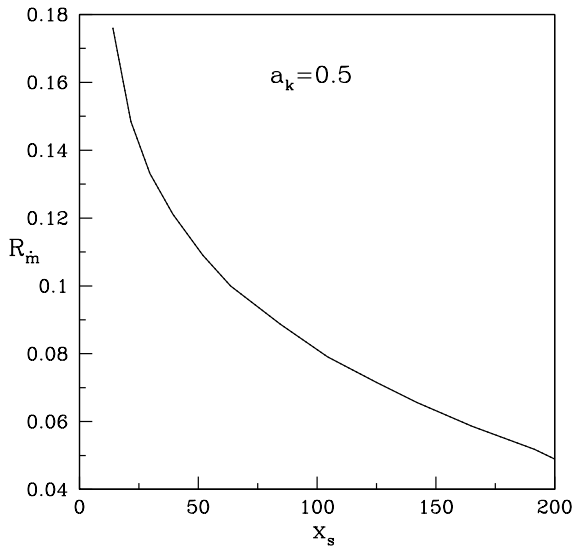}
\caption{Variation of outflow rate ($R_{\dot m}$) with shock location ($x_s$)
obtained for different sets of energy (${\mathcal E}$) and angular momentum ($\lambda$).
Here, $a_k=0.5$ is used for representation. See text for details.
}
\end{figure}
%%%%%%%%%%%%%%%%%%%%%%%%%%%%%%%%%%%%%%%%%%%%%%%%%%

Until now, we discuss the various properties of the accretion-ejection solutions
in terms of the inflow parameters. However, all these studies are based on
the assumption of stationary state that does not represent the dynamical
behaviour of accretion flows around the outbursting BH sources. Usually,
these sources change their accretion states with time and the characteristic
features of the emergent radiations and
the Jet properties are also varied accordingly. During these dynamical change of
states, it is unlikely that the dissipative properties (i.e., viscosity, cooling
processes etc.) of the accretion flow will remain constant all throughout, instead, they
seem to adjust in a such way that suitably represents the dynamical variation of PSC.
In {\it LHS} of GBHs, the typical geometry of PSC is quite large
which reduces considerably
in {\it HIMS} as the dynamical shock moves towards the black hole during the
rising phase of the outburst. Such a trend 
has been reported while modeling the evolution of QPO frequencies for several black hole sources 
\citep{Chakrabarti-etal08,Chakrabarti-etal09,Nandi-etal12,Debnath-etal13,Radhika-Nandi14,Iyer-etal15}.
Interestingly, it is also observed that most of the GBH sources show
persistent/steady radio emission during these accretion states \citep{Fender-etal04,Fender-etal09}.
Now, we attempt to address the evolution of accretion states from {\it LHS} to
{\it HIMS} and its association with mass loss using our present formalism as case by
case which possibly acts as a local model at individual time frame.
Toward this, a set of energy and angular momentum ($\mathcal{E}, \lambda$) of the
inflowing matter is identified which would represent their local values for a
dissipative accretion flow and obtain the shock location $x_s$ and outflow rate
$R_{\dot{m}}$ for a given $a_k$. Since the post-shock geometry (PSC) is associated with
$x_s$, as anticipated earlier, we chose various such sets of ($\mathcal{E}, \lambda$)
in a way that $x_s$ moves towards the black hole and show the variation of $R_{\dot{m}}$
with $x_s$ in Fig. 10 for a typical value of $a_k=0.5$.
Note that $R_{\dot{m}}$ is increased with the decrease of $x_s$
resulting more Jet kinetic power that represents the common characteristic of radio 
emissions observed in outbursting sources.

\section{Astrophysical Application}

Here, we attempt to estimate the mass outflow rate and its associated Jet kinetic
power for various astrophysical black hole sources (GBHs and AGNs) using our present
formalism. Since our work deals with the steady outflows, we focus only to those
sources particularly to their accretion states where persistent/steady Jets are
observed. These Jets are essentially compact (i.e., optically thick) in nature and
are not isolated completely from the core of the central engine
\citep{Mirabel-Rodriguez94,Mirabel-Rodriguez98,Corbel-etal00,Corbel-etal01,Fender-etal01}.

\subsection{Jet kinetic power predicted from Accretion States} 

We consider five black hole sources namely, XTE J1859+226, GRO J1655-40, GX 339-4,
H 1743-322 and GRS 1915+105, respectively. For these sources, mass ($M_{BH}$),
distance ($d$) and spin $(a_k)$ are constrained within the accuracy limit and are given
in Table 1. In order to estimate the Jet kinetic power, we calculate the X-ray
flux for these sources in the {\it low-hard state} ({\it LHS}) and
{\it hard-intermediate state} ({\it HIMS}) using {\it $RXTE^{1}$}
satellite data. 

In general, GBH sources undergo outbursts \citep{Homan-Belloni05,Remillard-McClintock06,
Debnath-etal08,Nandi-etal12,Debnath-etal13,Radhika-Nandi14,Iyer-etal15} and show 
activity of radio emissions coupled with the accretion states of their evolution during the 
outburst phases
\citep{Brocksopp-etal02,Fender-etal04,Fender-etal09,Cadolle-etal11,Millerjones-etal12,Radhika-Nandi14}.
It has also been observed that most of the outbursting sources show persistent/steady
Jet activity during {\it LHS} and {\it HIMS} while transient `relativistic' ejections
are observed during the transition from hard-intermediate to soft-intermediate state
\citep{Brocksopp-etal02,Fender-etal04,Fender-etal09,Radhika-Nandi14}.

In order to estimate the `model predicted' Jet power based on accretion states, we estimated
X-ray flux from a single observation of each states (i.e., {\it LHS} and {\it HIMS}) for
all the sources from different outburst phases (as mentioned in the Table-1). 
We use archival data obtained from the HEASARC database of the {\it RXTE} satellite for 
the estimation of X-ray flux for all the five sources. We extracted and analyzed background-subtracted
PCA (3 - 30 keV) spectral data using PCU2 detector (i.e., well calibrated detector) for our 
specific purpose. The standard FTOOLS package of HEASOFT v6.15.1 was used for spectral data reduction
(see \citet{Nandi-etal12,Radhika-Nandi14} for details). For spectral analysis and modeling,
we used the packages of XSPEC v12.8.1. 

We have done spectral modeling for each observations from each spectral 
states (i.e., {\it LHS} and {\it HIMS}) for all the five sources using PCA spectral data 
in the energy range of 3 - 30 keV. 
We model the energy spectrum using the phenomenological accretion disk model i.e., consisting of 
a {\it diskbb} and a {\it power-law} component. In this modeling, the {\it diskbb} and 
{\it power-law} components provide the contribution from the accretion
disk and `hot' Compton 
corona (i.e., PSC) and thereby one can estimate the total {\it unabsorbed} X-ray flux 
($F_x$, in units of ${\rm ergs~cm^{-2}~s^{-1}}$) emitted from the accretion disk around 
the GBH sources.

Once we obtain the {\it unabsorbed} X-ray flux ($F_x$) of a source, we calculate the X-ray luminosity ($L_x$) 
of the source employing the relation 
$L_x=4\pi d^2 F_x$, where $d$ is the distance of the source. 
Assuming the maximum radiative efficiency of the infalling matter around rotating black
hole is $\eta \sim 0.3$, we calculate the accretion rate of the black hole as,
$$
{\dot M}_{acc} = 2.99 \times 10^{-16}
\left(\frac{F_x~d^2} {c^2}\right)
\left( \frac{M_{BH}}{M_{\sun}}\right)^{-1}{\dot M}_{Edd}.
$$
Here, $M_{BH}$ denotes the mass of the black hole and ${\dot M}_{Edd}$
represents the Eddington accretion rate. Next, we calculate the mass outflow rate
using our theoretical estimate (Eq. 14) which has its maximum limit ($R^{\rm max}_{\dot m}$)
for a particular set of ($a_k, {\mathcal E}, \lambda$) (see Fig. 9). With this, we find the
{\it maximum mass outflow rate} as ${\dot M}_{out}=R^{\rm max}_{\dot m} {\dot M}_{acc}$ which
successively allows us to compute the {\it maximum Jet kinetic power} as,
$$
L^{\rm max}_{Jet} = R^{\rm max}_{\dot m} \times {\dot M}_{acc} \times c^2 ~{\rm ergs~s^{-1}}.
\eqno(15)
$$

As pointed out earlier that steady and persistent radio emissions are observed during
{\it LHS} and {\it HIMS} \citep{Fender-etal04,Fender-etal09}. In addition, there are
indications that the radio emission increases while the state transition from {\it LHS}
to {\it HIMS} takes place. Moreover, the radio emission becomes non-steady during the
{\it HIMS} itself just prior to the ejection and it is quite known
that the relativistic
ejections take place during the transition from {\it HIMS} to {\it SIMS}
\citep{Fender-etal04,Fender-etal09}. All these observations perhaps indicate that the
persistent Jet activity would be maximum during {\it HIMS}. Therefore, in order to
study the outflow properties in the {\it HIMS}, we consider the maximum outflow rate
obtained from our model calculation (see Fig. 9) and use it to estimate the Jet kinetic power.
On the other hand, Jet kinetic power in the {\it LHS} is expected to be lower than
the {\it HIMS} although its quantitative estimate is unclear.
Therefore, we use outflow rate to be
around $\sim 10\%$ as a representative value while calculating the Jet kinetic
power for {\it LHS}. Below, we mention the fundamental properties of each sources and present the
details for the estimation of Jet kinetic power corresponding to individual states of
a particular observation.

%%%%%%%%%%%%%%%%%%%%%%%%%%%%%%%%%%%%%%%%%%%%%%%%%%
\begin{center}
\begin{table*}
\begin{minipage}{180mm}
\caption{\label{tabflares} Accretion state dependent Jet kinetic power.}
 \label{symbols}
 \begin{tabular}{lccccccccc}
  \hline
   Objects & $M_{BH}$     & $d$   & $a_k$ & \multicolumn{2}{c}{Observation ($3-30~{\rm keV}$)} 
   & $\dot M_{acc}$ & $R_{\dot m}$ & $L_{Jet}$ & $L_{Jet}^{Obs}$~$^\#$\\\\
   \cline{5-6}\\
   &  ($M_{\sun}$) & (kpc) &  &  States & $F_x$~(${\rm ergs~cm^{-2}~s^{-1}}$) & (${\dot M}_{Edd}$)$^{*}$
   & (\%) & (${\rm ergs~s^{-1}}$) & (${\rm ergs~s^{-1}}$)\\
   \hline
   XTE J1859+226 & 7 & 11 & 0.4 & $LHS$ & $5.71 \times 10^{-9}$ & $0.304$ & 9.83  & $2.52 \times 10^{37}$ &\\
  (2009 Outburst) &  &  & & $HIMS$ & $12.79 \times 10^{-9}$ & $0.680$ & 17.5 & $1.08 \times 10^{38}$ &\\
   & & & & & & & & & $1.82 \times 10^{38}$ (1)\\
   & & & & & & & & &\\
   GRO J1655-40 & 6.3 & 3.2 & 0.7 & $LHS$ & $3.19\times 10^{-9}$ & $0.016$ & 9.98  & $1.18 \times 10^{36}$ &\\
   (2005 Outburst) & & & & $HIMS $& $8.02 \times 10^{-9}$ & $0.040$ & 17.68 & $5.79 \times 10^{36}$ &\\
   %& & & & & & & & & \textcolor{red}{$8.19 \times 10^{38}$ (1)}\\
   & & & & & & & & & $3.01 \times 10^{36}$ (2)\\
   & & & & & & & & &\\
   GX 339-4 & 7.5 & 15 & 0.4 & $LHS$ & $15.71\times 10^{-9}$ & $1.450$ & 9.98 & $1.29 \times 10^{38}$ &\\
 (2002 Outburst)   &  &  &  & $HIMS^{**}$ & $12.12 \times 10^{-9}$ & $1.118$ & 17.5 & $1.90 \times 10^{38}$ &\\
   & & & & & & & & & $2.92 \times 10^{38}$ (1)\\
   & & & & & & & & &\\
   H 1743-322 & 8 & 8.5 & 0.2 & $HIMS$ & $4.03 \times 10^{-9}$ & $0.112$ & 17.35 & $2.01 \times 10^{37}$ &\\
  (2009 Outburst)  & & & & & & & & & $1.08 \times 10^{38}$ (3)\\
   %& & & & & & & & & \textcolor{red}{$1.08 \times 10^{38}$ (3)}\\
   & & & & & & & & &\\
   GRS 1915+105 & 12.4 & 8.6 & $> 0.98^{\dagger}$ & $LHS$ & $20.33\times 10^{-9}$ & $0.373$ & 9.98
   & $5.99\times 10^{37}$ &\\
  (1997 Observation)  & & & & & & & & & $8.06 \times 10^{37}$ (1)\\
\\ 
  \hline
 \end{tabular}

$^{*}~{\dot M}_{Edd}=1.44\times 10^{17}\left(\frac{M_{BH}}{M_{\odot}}\right){\rm gm~{s}^{-1}}$\\
$^{**}$~Total flux in {\it HIMS} ($3-30~{\rm keV}$) is smaller than {\it LHS} as the
contribution of the hard X-ray flux ($> 10~{\rm keV}$) in {\it HIMS} is %significantly
less. \\
$^{\#}~$$L_{Jet}^{Obs}=\eta{\dot M}_{\rm out} c^2$ is used for the source $H~1743-322$,
where ${\dot M}_{\rm out}$ is the outflow rate (see reference). For other sources,
$L_{Jet}^{Obs}=L_{r} \times L_{Edd}$ is used, where $L_{r}$ is the observed Jet power (in $Edd$)
(see references) and $L_{Edd}=1.3\times 10^{38}\left(\frac{M_{BH}}{M_{\odot}}\right){\rm ergs~{s}^{-1}}$.\\ 
References: ($1$) \citet{Fender-etal04} ($2$) \citet{Migliari-etal07} ($3$) \citet{Miller-etal12}\\

\end{minipage}
\end{table*}
\end{center}
%*****************************************************

\vskip 1.0cm
\noindent{XTE J1859 + 226:}

\citet{Filippenko-Chornock01} first presented the dynamical estimate of mass of the 
source to be around $7.4 \pm 1.1 M_{\sun}$. 
Recently, \citet{Radhika-Nandi14} claimed that the mass of XTE J1859 + 226
is perhaps in between $6.58M_{\sun}-8.84M_{\sun}$ which is similar to the prediction
of \citet{Shaposhnikov-Titarchuk09} although the lower mass limit is estimated
as $5.4M_{\sun}$ by \citet{Corral-santana-etal01}. However, we consider the typical 
mass of the source as $7M_{\sun}$. The distance of this source
is around $d \sim 11 {\rm kpc}$ \citep{Filippenko-Chornock01}. \citet{Steiner-etal13}
measured the spin as $a_k\sim 0.4$, however, \citet{Motta-etal14a} recently
reported the spin of the source is $a_k\sim 0.34$. Since the spin predictions
are quite close, we use $a_k\sim 0.4$ for this analysis. We
estimate the fluxes $F_x$ (see Table 1) of
{\it LHS} and {\it HIMS} of the 2009 outburst of the source \citep{Radhika-Nandi14}.
The corresponding disc luminosities are calculated as 
$L_{\rm disc}^{LHS}=8.26 \times 10^{37}~{\rm ergs~s^{-1}}$
and $L_{\rm disc}^{HIMS}=1.85 \times 10^{38}~{\rm ergs~s^{-1}}$, respectively.
Now, it is reasonable to assume the accretion efficiency for rotating black hole
as $\eta = 0.3$ which corresponds to the accretion rate of the inflowing matter as 
${\dot M}_{acc}^{LHS} = 0.304 {\dot M}_{Edd}$ 
in {\it LHS} and 
${\dot M}_{acc}^{HIMS} = 0.680 {\dot M}_{Edd}$ 
in {\it HIMS}. 
For $LHS$, we use $R_{\dot m}=9.83\%$ following our theoretical estimate where
$x_s=64.6r_g$ for $a_k=0.4$, ${\mathcal E}=0.00198$ and $\lambda=3.18$. Incorporating
these inputs in Eq. (15), we obtain the Jet kinetic power as
$L^{LHS}_{Jet} = 2.52\times 10^{37}~{\rm ergs~s^{-1}}$. The maximum mass
outflow rate for {\it HIMS} corresponding to  $a_k=0.4$ is obtain from Fig. 9 as
$R^{\rm max}_{\dot m}=17.5\%$ for ${\mathcal E}=0.00547$, and $\lambda=3.1$, where
the shock transition occur at $21.9r_g$. Using these values in Eq. (15), we obtain the
maximum Jet kinetic power as $L^{HIMS}_{Jet} = 1.08\times 10^{38}~{\rm ergs~s^{-1}}$ which we
regard to be associated with the {\it Hard-Intermediate} state of this source.

\vskip 0.5cm
\noindent{GRO J1655-40:}

The mass of the source GRO J1655-40 is reported by \citet{Greene-etal01} and is given
by $6.3M_{\sun}$. Recently, \citet{Motta-etal14b} estimated the object mass as 
$5.31 \pm 0.07~{\rm M}_{\sun}$.
This source is located at around $d\sim 3.2~{\rm kpc}$
and the spin of this source is estimated by \citet{Shafee-etal06} using {\it RXTE$^1$}
and {\it ASCA} data through the modeling of the thermal spectral continuum and obtained
in the range $a_k\sim 0.65-0.75$ (and reference therein).
Fitting the strong reflection
features of iron line in {\it XMM-Newton} data, \citet{Reis-etal09} determines the lower
limit of the spin of this object $a_k=0.9$. \citet{Motta-etal14b} calculated
the spin of the object using X-ray timing method and found as $a_k = 0.290 \pm 0.003$ which
is an inconsistent estimate compared to iron line or continuum methods.
In our present analysis, however, we consider $a_k=0.7$.
As before, we analyzed the 2005 outburst of GRO J1655-40
to calculate the X-ray fluxes ($F_x$) for {\it LHS} and {\it HIMS} (see Table 1)
in the energy range $3-30~{\rm keV}$ and obtain the corresponding accretion rates 
${\dot M}_{acc}^{LHS} = 0.016 {\dot M}_{Edd}$ in {\it LHS} and 
${\dot M}_{acc}^{HIMS} = 0.040 {\dot M}_{Edd}$ in {\it HIMS}.
We use $R_{\dot m}=9.98\%$ in case of $LHS$ which is obtained
for $a_k=0.7$, ${\mathcal E}=0.00258$ and $\lambda=2.845$ where $x_s=49.98r_g$. From these
values, we estimate Jet kinetic power in $LHS$ as 
$L^{LHS}_{Jet} = 1.18\times 10^{36}~{\rm ergs~s^{-1}}$.
For $HIMS$, we estimate of the maximum mass outflow rate for $a_k=0.7$
(see Fig. 9) as $R^{\rm max}_{\dot m}=17.68\%$ where ${\mathcal E} = 0.0073$
and $\lambda = 2.715$ with $x_s=16.73r_g$. These inputs provide the maximum Jet kinetic power
as $L^{HIMS}_{Jet} = 5.79\times 10^{36}~{\rm ergs~s^{-1}}$.

\vskip 0.5cm
\noindent{GX 339-4:}

The mass of the object is estimated as $7.5 \pm 0.8~{\rm M}_{\sun}$ \citet{Chen11} and
distance $d \sim 15~{\rm kpc}$ by \citet{Hynes-etal04}. The issue of spin measurement of this
objects is not settled yet as there are conflicting measurements. Analyzing the 
{\it XMM-Newton} data set for broad iron line detection, \citet{Reis-etal08} and
\citet{Miller-etal08}
claimed $a_k=0.935$ while the disc continuum fitting prefers the lower spin having upper
limit of  $a_k < 0.9$ \citep{Kolehmainen-Done10}. However, analyzing the wide-band
{\it Suzaku} speatra, \citet{Yamada-etal09} pointed out $a_k < 0.4$.
Being aware of these, we consider a conservative estimate of spin as $a_k = 0.4$ in
our calculation. This source has undergone outburst phases several times during RXTE era.
For this analysis, we consider the 2002 outburst spectral data (in the energy band of
$3-30$ keV) for {\it LHS} and {\it HIMS}. 
During May 2 of 2002 outburst, the object was in {\it LHS} emitting X-ray flux of
$F^{LHS}_{x}=15.71\times10^{-9}~{\rm ergs~cm^{-2}~s^{-1}}$ (see Table 1) and the disc
luminosity of $L^{LHS}_{disc}=4.22\times 10^{38}~{\rm ergs~s^{-1}}$. Using $\eta=0.3$,
the disc rate is calculated as ${\dot M}_{acc}^{LHS} = 1.450 {\dot M}_{Edd}$.
Estimating $R_{\dot m}=9.98\%$ from our model using  ${\mathcal E} = 0.00198$ and
$\lambda = 3.18$ with $x_s=64.63r_g$, we calculate the Jet kinetic power
$L^{LHS}_{Jet} = 1.29\times 10^{38}~{\rm ergs~s^{-1}}$. On May 12 of the 2002 outburst,
the source was in {\it HIMS} and the radiated X-ray flux was calculated as
$F^{HIMS}_{x}=12.12\times10^{-9}~{\rm ergs~cm^{-2}~s^{-1}}$ (see Table 1) and the disc
luminosity of $L^{HIMS}_{disc}=3.26\times 10^{38}~{\rm ergs~s^{-1}}$. Using $\eta=0.3$,
the disc rate is calculated as ${\dot M}_{acc}^{HIMS} = 1.118 {\dot M}_{Edd}$.
Estimating $R^{max}_{\dot m}=17.5\%$ from our model using  ${\mathcal E} = 0.00547$ and
$\lambda = 3.1$ with $x_s=21.9r_g$, we calculate the maximum Jet kinetic power
$L^{HIMS}_{Jet} = 1.90\times 10^{38}~{\rm ergs~s^{-1}}$.

\vskip 0.5cm
\noindent{H 1743-322:}

\citet{Miller-etal12} and \citet{Steiner-etal12}
reported the mass, distance and spin of H 1743-322 as 
$\sim 8M_{\sun}$, $d\sim 8.5~{\rm kpc}$ and $a_k=0.2$, respectively. Using the
RXTE observation of the 2009 outburst of the source, the disc luminosity for this source is computed for
the energy range of $3-30${\rm ~keV} corresponding to the measured X-ray flux of
$F^{HIMS}_{x}=4.03\times10^{-9}~{\rm ergs~cm^{-2}~s^{-1}}$ in {\it HIMS} (see Table-1) 
\citep{Millerjones-etal12} as $4.63\times 10^{37}~{\rm ergs~s^{-1}}$.  Assuming $\eta = 0.3$,
we obtain the accretion rate as ${\dot M}_{acc}^{HIMS} = 0.112 {\dot M}_{Edd}$.
Accretion rate in {\it LHS} is not known due to lack of observation during
the 2009 outburst \citep{Millerjones-etal12}. 
For $a_k=0.2$, ${\mathcal E} =0.00485$ and $\lambda=3.176$, we obtain the maximum mass outflow rate 
for {\it HIMS} as $R^{\rm max}_{\dot m}=17.35\%$ with $x_s=23.66r_g$. Employing these
values, we find the maximum Jet kinetic power as $L^{HIMS}_{Jet} = 2.01\times 10^{37}~{\rm ergs~s^{-1}}$.

\vskip 0.5cm
\noindent{GRS 1915+105:}

Earlier, \citet{Greiner-etal01} estimated the mass of GRS 1915+105 as $(14 \pm 4)~M_{\sun}$.
Recently, \citet{Hurley-etal13} reported the new mass estimate as $(12.9 \pm 2.4)~M_{\sun}$
which is further revised as $(12.4 \pm 2)~M_{\sun}$ by \citet{Reid-etal2014}. The distance
of the source is reported as $d \sim (9.4 \pm 0.2)$ kpc by \citet{Hurley-etal13} and \citet{Reid-etal2014}
claimed the distance to be $d \sim (8.6 \pm 2)$ kpc. However, in this calculation we use the mass
and distance of the source as $12.4 M_{\sun}$ and $8.6$ kpc, respectively in order to compute the disc
luminosity. The source is extremely rotating as \citet{McClintock-etal06} estimated the spin
parameter $a_k > 0.98$ which is similar to the estimate of \citet{Blum-etal09}.
As the source is highly variable in X-rays, we choose one observation when the source
was in the {\it hard state} that is similar to {\it LHS} of other GBH sources. {\it RXTE}
observed the source in the energy range of $3-30$ keV on $22^{th}$ October, 1997 and object was
in so-called $\chi$-class \citep{Belloni-etal01,Nandi-etal01c}.
For this observation, the disc luminosity corresponding to the X-ray flux of 
$F^{LHS}_{x}=20.33\times10^{-9}~{\rm ergs~cm^{-2}~s^{-1}}$ (see Table 1)
is obtained as $1.79\times 10^{38}~{\rm ergs~s^{-1}}$ which provides the
disc accretion rate as ${\dot M}_{in}^{LHS} = 0.373 {\dot M}_{Edd}$. According to
our model, the mass outflow rate for {\it LHS} (or simply hard state for this source) is
calculated as $R_{\dot m}=9.98\%$ where, $x_s=50.47r_g$ with $a_k=0.98$,
${\mathcal E} = 0.00276$ and $\lambda=2.679$. Here, we cross the upper limit of $a_k$
(indicated by dagger ($^\dagger$) in column 4 of Table 1) as the adopted black hole potential
satisfactorily describes the space-time geometry for $a_k\lesssim 0.8$. However, we anticipate
that the obtained $R_{\dot m}$ provides qualitative estimate that would not differ
significantly from its exact value. Using these values,
the Jet kinetic power is found to be $L^{LHS}_{Jet} = 5.99\times 10^{37}~{\rm ergs~s^{-1}}$.

In this Section, we calculated the Jet kinetic power mostly for outbursting GBH sources (except GRS 1915+105)
for different accretion states (i.e., {\it LHS} and {\it HIMS}). Our findings clearly show that as
the sources transit from {\it LHS} to {\it HIMS}, there is significant increase in the
Jet kinetic power. The predicted Jet kinetic powers are in close agreement with the observed values 
for XTE J1859+226, GRO J1655-40, GX 339-4, GRS 1915+105 and H 1743-322
\citep{Fender-etal04,Migliari-etal07,Miller-etal12}.

\subsection{Jet kinetic power estimated from Accretion Rates} 

%%%%%%%%%%%%%%%%%%%%%%%%%%%%%%%%%%%%%%%%%%%%%%%%%%
%%%%%%%%%%% TABLE %%%%%%%%%%%%%%%%%%%%%%%%%%%%%%%%
%%%%%%%%%%%%%%%%%%%%%%%%%%%%%%%%%%%%%%%%%%%%%%%%%%
%choton
\begin{center}
\begin{table*}
\begin{minipage}{180mm}
 \caption{Estimated Jet kinetic power from accretion rate.}
 \label{symbols}
 \begin{tabular}{@{}lccccccccc}
  \hline
  Objects & $M_{BH}$ & $\dot M_{acc}$ & $a_k$ & ${\mathcal E}$ & $\lambda$ & $x_s$ & $R_{\dot m}^{\rm max}$ & 
  $L^{\rm max}_{Jet}$ & $L_{Jet}^{Obs}$~$^\#$\\
  & ($M_{\sun}$) & (${\dot M}_{Edd}$) &  & ($c^2$) & ($cr_g$) & ($r_g$) & (\%) & (${\rm ergs~s^{-1}}$) & 
  (${\rm ergs~s^{-1}}$) \\
  \hline
   A0620-00 & $6.60~(a)$  & $1.684~(b)$ & $0.12~(b)$ & 0.00444 & 3.25 & 26.84 & 17.25 & $2.48\times10^{38}$ 
   & ...\\
   & & & & & & & & \\
   LMC X-3 & $6.98~(c)$ & $2.487~(d)$ & $0.25~(e)$ & 0.00476 & 3.15 & 25.72 & 17.35 & $3.90\times10^{38}$ 
   &...\\
   & & & & & & & & \\
   XTE J1550-564  & $9.10 ~(f)$ & $0.511~(g)$ & $0.34 {~(g)}$ & 0.00530 & 3.06 & 22.07 & 17.43 & $1.05\times10^{38}$ 
   & $3.55 \times 10^{38}~(y)$\\ 
   & & & & & & & & \\
   M33 X-7 & $15.66~(h)$ & $0.718~(i)$ & $0.84^{\dagger}~(j)$ & 0.00807 & 2.61 & 15.60 & 17.79 & $2.59\times10^{38}$
   &...\\
   & & & & & & & & \\
   4U 1543-47 & $9.40~(k)$ & $1.315~(l)$ & $0.43~(l)$ & 0.00565 & 2.98 & 21.07 & 17.48 & $2.80\times10^{37}$ 
   &...\\
   & & & & & & & & \\
   LMC X-1 & $10.90~(m)$ & $0.853~(n)$ & $0.92^{\dagger}~(n)$ & 0.00830 & 2.58 & 15.42 & 17.93 & $2.16\times10^{37}$
   &...\\
   & & & & & & & & \\
   Cyg X-1 & $14.80~(o)$ & $0.061~(p)$ & $\ge 0.95^{\dagger}~(p)$ & 0.00835 & 2.58 & 15.38 & 17.98 & $2.10\times10^{37}$
   & $3.85 \times 10^{37}~(y)$\\
   & & & & & & & & \\
   Mrk 79 & $5.24 \times 10^7~(q)$ & $9.82\times 10^{-2}~(r)$ & $0.70~(s)$ & 0.00730 & 2.72 & 16.73 & 17.68 & $1.18\times10^{44}$
   &...\\
   & & & & & & & & \\
   M87 & $3.50 \times 10^9~(t)$ & $1.16 \times 10^{-4}~(u)$ & $\ge0.65~(v)$ & 0.00705 & 2.76 & 16.94 & 17.62 & $9.25\times10^{42}$
   & $1.00 \times 10^{45}~(z)$\\
   & & & & & & & & \\
   Sgr A* & $4.90 \times 10^6~(w)$ & $7.89 \times 10^{-5}~(x)$ & $0.99^{\dagger}~(w)$ & 0.00859 & 2.57 & 14.71 & 18.09 
   & $9.07\times10^{39}$ & $1.00 \times 10^{39}~(zz)$\\
  \hline
 \end{tabular}
%\\
$^{\#}~L_{Jet}^{Obs}=L_{r} \times L_{Edd}$, where $L_{r}$ is the observed Jet power (in $Edd$) (see references).\\
References: ($a$) \citet{Cantrell-etal10} ($b$) \citet{Gou-etal10} ($c$) \citet{Orosz-etal14} ($d$) \citet{Kubota-etal10}
($e$) \citet{Steiner-etal14} ($f$) \citet{Orosz-etal11a} ($g$) \citet{Steiner-etal11} ($h$) \citet{Orosz-etal07}
($i$) \citet{Liu-etal08} ($j$) \citet{Liu-etal10} ($k$) \citet{Orosz-03} ($l$) \citet{Morningstar-Miller14}
($m$) \citet{Orosz-etal09} ($n$) \citet{Gou-etal09} ($o$) \citet{Orosz-etal11b} ($p$) \citet{Gou-etal11}
($q$) \citet{Peterson-etal04} ($r$) \citet{Riffel-etal13} ($s$) \citet{Gallo-etal11} ($t$) \citet{Walsh-etal2013}
($u$) \citet{Kuo-etal14} ($v$) \citet{Wang-etal08} ($w$) \citet{Aschenbach10} ($x$) \citet{Yuan-etal02}
($y$) \citet{Fender-etal04} ($z$) \citet{deGasperin-etal12} ($zz$) \citet{Falcke-Biermann99}

\end{minipage}
\end{table*}
\end{center}

We further extend our study to estimate the Jet kinetic power for other
GBH and AGN sources where we do not observe fast ($\sim$ day scale)
state transitions similar to outbursting GBH sources (except XTE J1550-564).
In order to calculate the Jet kinetic
power for these sources, we find their mass ($M_{BH}$), accretion rate
(${\dot M}_{acc}$) and spin ($a_k$) from the existing literature.
Meanwhile, we compute $R^{\rm max}_{\dot m}$ corresponding to black hole
spin $a_k$ using our theoretical approach. We then use the values of
${\dot M}_{acc}$ and $R^{\rm max}_{\dot m}$ in Eq. (15) to estimate the 
Jet kinetic power.

In Table 2, we display the physical parameters of the sources (GBHs and AGNs)
under consideration along with the computed Jet kinetic power obtained from
our analysis.  In column 1-4, we present the list of sources, their mass
($M_{BH}$), accretion rate (${\dot M}_{in}$) and spin ($a_k$). In column 5-6,
we mention the representative values of energy $\mathcal E$,
angular momentum $\lambda$ of the inflow that provide the shock location
$x_s$ (in column 7) and the corresponding {\it maximum mass outflow rate} $R^{\rm max}_{\dot m}$
(in column 8). Finally, in column 9, we present the maximum Jet kinetic power 
$L^{\rm max}_{Jet}$.  The first seven sources are GBHs whereas the last three sources are AGNs.
Here, we give emphasis on
the maximum $R_{\dot m}$ in order to speculate the upper limit of Jet kinetic power.
We find that the estimated Jet kinetic powers are in close agreement with the
observed values at least for few sources, namely  Cyg X-1, XTE J1550-564, M87 and Sgr A*
\citep{Fender-etal04,deGasperin-etal12,Falcke-Biermann99}. For remaining sources,
we argue that the present method illustrates the typical
estimates of Jet kinetic power that possibly lie within the acceptable range.
Further, in our analysis, we choose four sources having black hole spin $a_k>0.8$ which
are indicated by dagger ($^\dagger$) in column 4. As before, we anticipate that
the obtained $L^{\rm max}_{Jet}$ for these sources provide qualitative estimates that
would not differ significantly from their observed values. 

\section{Concluding remarks}

In this work, we self-consistently examine the accretion-ejection mechanism around the 
rotating black holes. We consider the structure of the accretion disc to be
stationary, thin, rotating and advection dominated which contains R-H shock waves.
These shocks are formed as a consequence of centrifugal barrier located near the
black hole horizon. Based on the second law of thermodynamics, we argue that
the shock solutions are dynamically preferred over the smooth solutions as they
possess higher entropy (Becker \& Kazanas 2001).
During accretion, a part of the super-sonic matter is deflected
at the centrifugal barrier due to excess thermal pressure caused by the shock compression
and eventually emerge out in the form of thermally driven outflows. These outflows
are further channeled through the confined geometry bounded by the funnel wall 
and pressure maxima surface along the rotation axis of the black hole \citep{Molteni-etal96}.

We find that mass loss can occur for prograde as well as retrograde flows. When outflow
is emerged out from the inner part of the disk (PSC), post-shock pressure is decreased
that essentially compels the shock front to move forward towards the horizon 
in order to maintain the pressure balance across the shock. Therefore, a flow originally
containing shock wave close to its minimum location ($x^{\rm min}_s$) for $R_{\dot m}=0$,
will not provide any outflow as R-H shock conditions are not favorable there
(see Fig. 3).  This certainly tells us that the outflow solutions would be restricted compared
to the global shocked
accretion solutions in absence of mass loss. However, we show that the
shock induced global inflow-outflow solutions are not isolated solutions, but exist
for a wide range of inflow parameters, namely, $\mathcal E$ and $\lambda$, respectively. 
Interestingly, numerical simulations also indicate the similar findings as reported by
\citet{Das-etal14}. We examine the existence of such solutions with and without
outflow and obtain the parameter space spanned by the $\mathcal E$ and $\lambda$
as function of black hole rotation parameter $a_k$. As anticipated above, the parameter
space that provides mass outflow is shrunk compared to the case of no outflow
(see Fig. 7). In other words, a significant region of the parameter space that
exhibits stationary
shock for $R_{\dot m}=0$ does not provide steady shock solutions in presence of
outflow. This possibly infer that the accreting matter having $\mathcal E$
and $\lambda$ from this region of the parameter space may demonstrate the
non-steady behavior \citep{Das-etal01b,Das07} 
which is triggered simply due to the presence of mass loss.
Investigation of such scenario requires time-dependent calculations which 
is beyond the scope of the present work and we wish to report this elsewhere. 

In this work, we mainly focused on thermally
ultra-relativistic flows with adiabatic index $\gamma = 4/3$. However, when the
cooling effects are negligible, the purely non-relativistic flow behaves like
gas-pressure dominated with $\gamma \sim5/3$. In reality, the value of $\gamma$
would be an intermediate value depending on the dissipation processes active in
the flow. Keeping this in mind, we calculate the parameter space for different
values of $\gamma$ and find that shock forms in all the cases even in presence of
outflow. Interestingly, we observe that the effective region of the parameter
space is reduced with the increase of $\gamma$ indicating the limited possibility
of shock transition when the flow changes its character towards the non-relativistic
regime.

We have estimated the mass outflow rate using the inflow parameters, such as, $\mathcal E$
and $\lambda$. We show that $R_{\dot m}$ increases with $a_k$ for a fixed $\mathcal E$
and $\lambda$ (see Fig. 4). This is possibly due to the fact that as $a_k$ is increased shock
forms away from the black hole. Therefore, the effective area of the post-shock flow
(PSC), where the inflowing matter deflects to generate outflow, becomes large and results
enhanced outflow rate. 
Further, we attempt to find the maximum mass loss ($R^{\rm max}_{\dot m}$)
from the disc and quantify it in terms of the inflow parameters. For this, we
fix the value of $\gamma$ and explore all possible combinations
of energy and angular momentum to obtain $R^{\rm max}_{\dot m}$. In Fig. 9, we
show the variation of $R^{\rm max}_{\dot m}$ with $a_k$ and observe that flow with
$\gamma=4/3$ exhibits the highest outflow rate $R^{\rm max}_{\dot m}$ that lies
in the range around $\sim 17\%-18\%$. Also, very weak correlation
is seen between $R^{\rm max}_{\dot m}$ and $a_k$ in all the cases.
Moreover, we have shown in Fig. 10 that for various sets of (${\mathcal E}, \lambda$),
$R_{\dot m}$ increases with the decrease of $x_s$ (equivalently Jet power increases
as the size of the PSC is reduced) based on the realistic scenario (i.e., 
spectral states transit from
{\it LHS} to {\it HIMS}) which seems to be a common characteristic observed in GBH sources.

We employ our formalism in order to estimate the Jet kinetic power ($L_{Jet}$) of
several black hole sources (GBHs and AGNs). To begin with, 
we consider outbursting sources to calculate
$L_{Jet}$ based on the accretion states. For these sources, we compute their {\it unabsorbed}
X-ray fluxes in {\it LHS} and {\it HIMS} using the data of {\it RXTE}
observation and obtain their disc rate considering the accretion efficiency
$\eta = 0.3$ which seems to be relevant for rotating black holes. Then, we
calculate the outflow rate using our formalism and employ it in Eq. (15) along with the
disc rate to obtain $L_{Jet}$. While estimating the Jet kinetic power, we consider maximum
outflow rate $R^{\rm max}_{\dot m}$ for {\it HIMS} (see Fig. 9) and $R_{\dot m}\sim 10\%$
for {\it LHS}, as discussed in Section 5. In Table 1, we summarize the physical
parameters of five black hole sources along with the computed Jet kinetic power.
In the process of estimating $L^{\rm max}_{Jet}$, two quantities play major role.
First one is the unabsorbed X-ray flux computed from {\it RXTE} archival data
and the other is the maximum outflow rate obtained from our model calculation.
We find that the estimated $L^{\rm max}_{Jet}$ for various sources are in close agreement with
the observed values \citep{Fender-etal04,Miller-etal12}.
We further extend our study for other black hole sources where the fast
accretion state transitions ($\sim$ day scale) are not seen (except XTE J1550-564). 
For these sources, the disc rate is obtained from the existing literature (see Table 2)
and $R^{\rm max}_{\dot m}$ is computed from our theoretical calculation which finally 
provides the $L^{\rm max}_{Jet}$. We present these results in Table 2.
We notice that estimated $L^{\rm max}_{Jet}$ for Cyg X-1, XTE J1550-564, M87 and Sgr A*
are also in close agreement with the observed values 
\citep{Fender-etal04,deGasperin-etal12,Falcke-Biermann99}.
Following this findings, we argue that
$L^{\rm max}_{Jet}$ for rest of the sources would also in turn render the representative values
which are expected to be consistent with their actual estimates.

The present work has limitations as it is developed based on some approximations. We use
pseudo-Kerr gravitational
potential to mimic the general relativistic effect around rotating black hole that allows us to examine
the properties of non-linear shock solutions in presence of mass loss in a simpler way. We
ignore the effect of viscosity and radiative processes. We consider constant adiabatic index
instead of calculating it self-consistently based on its thermal properties. 
Needless to mention that we have not address the issue of transient relativistic ejections
and its collimation mechanism as we consider the Jet geometry only up to its sonic point.
Although the implementation of all such issues are beyond the scope of this paper, however,
we believe that the above approximations will not alter our basic conclusions qualitatively.

\section*{Acknowledgments}
We thank the reviewer for suggestions and comments that help us to
improve the manuscript. We would like to thank Indranil Chattopadhyay and Samir Mandal for
discussions.
This research has made use of the data obtained through High Energy Astrophysics Science Archive
Research Center on-line service, provided by NASA/Goddard Space Flight Center. 
AN acknowledges Group Director, SAG, Deputy Director, CDA and Director, ISAC,
for continuous support to carry out this research at ISAC, Bangalore.

 \label{lastpage}

\end{document}